\documentclass[aip,graphicx]{revtex4-1}
\usepackage[T1]{fontenc}
\usepackage[latin1]{inputenc}
\usepackage{graphicx}
\usepackage{setspace}
\usepackage[usenames,dvipsnames]{color}
\usepackage{amsmath}
\usepackage{MnSymbol}
\usepackage{wasysym}

\makeatletter

\newcommand{\TAE}{TAE}
\newcommand{\TAEs}{TAEs}
\newcommand{\MHD}{MHD}

\newcommand{\CKA}{CKA}
\newcommand{\VMEC}{VMEC}
\newcommand{\WSX}{W7-X}
\newcommand{\EAE}{EAE}
\newcommand{\EAEs}{EAEs}
\newcommand{\CHS}{CHS}
\newcommand{\HOneNF}{H-1NF}
\newcommand{\NGAE}{NGAE}

\makeatother
\begin{document}

\title
{Calculation of continuum damping of Alfv\'{e}n eigenmodes in 2D and 3D cases}

\author{G.W. Bowden}
\affiliation{Plasma Theory and Modelling, Research School of Physics and Engineering, Australian National University, Acton 2601, ACT Australia}
\author{M.J.~Hole}
\affiliation{Plasma Theory and Modelling, Research School of Physics and Engineering, Australian National University, Acton 2601, ACT Australia}
\author{A.~K{\"{o}}nies}
\affiliation{Max-Planck-Institut f\"{u}r Plasmaphysik, EURATOM-Association, D-17491 Greifswald, Germany}

\begin{abstract}
In ideal \MHD , shear Alfv\'{e}n eigenmodes may experience dissipationless damping due to resonant interaction with the shear Alfv\'{e}n continuum. This continuum damping can make a significant contribution to the overall growth/decay rate of shear Alfv\'{e}n eigenmodes, with consequent implications for fast ion transport. One method for calculating continuum damping is to solve the \MHD\ eigenvalue problem over a suitable contour in the complex plane, thereby satisfying the causality condition. Such an approach can be implemented in three-dimensional ideal \MHD\ codes which use the Galerkin method. Analytic functions can be fitted to numerical data for equilibrium quantities in order to determine the value of these quantities along the complex contour. This approach requires less resolution than the established technique of calculating damping as resistivity vanishes and is thus more computationally efficient. The complex contour method has been applied to the three-dimensional finite element ideal \MHD\ code \CKA . In this paper we discuss the application of the complex contour technique to calculate the continuum damping of global modes in tokamak as well as torsatron, \WSX\ and \HOneNF\ stellarator cases. To the authors' knowledge these stellarator calculations represent the first calculation of continuum damping for eigenmodes in fully three-dimensional equilibria. The continuum damping of global modes in \WSX\ and \HOneNF\ stellarator configurations investigated is found to depend sensitively on coupling to numerous poloidal and toroidal harmonics.
\end{abstract}

\pacs{52.30.Cv, 52.35.Bj, 52.55.Fa, 52.55.Hc, 52.65.Kj}

\maketitle

\section{Introduction}
Shear Alfv\'{e}n modes are a well-established feature of toroidal magnetically confined plasmas. These weakly damped modes are of substantial interest to fusion research as they may become unstable through resonant interaction with fast particles. This can degrade fast particle confinement, reducing plasma heating and damaging plasma facing components. Inhomogenous toroidal plasmas give rise to a continuum of modes of varying frequency, resonant on particular flux surfaces. Asymmetries in the magnetic geometry of the plasma lead to couplings between different poloidal and toroidal harmonics, resulting in gaps in the continuous spectrum in which global modes can occur. Examples of this type of gap mode include the toroidicity and elipticity induced shear Alfv\'{e}n eigenmodes (the \TAE\ and \EAE\ respectively) \cite{low_n_shear_Alfven_spectra,stability_of_Alfven_gap_modes_in_burning_plasmas}. Global modes can also occur due to extrema of the continuous spectrum, which can arise from variation of the magnetic geometry and density of the plasma \cite{excitation_of_global_eigenmodes_of_the_Alfven_wave_in_tokamaks}.

Resonant power absorption occurs where the eigenfrequency coincides with the frequency of a continuum branch. Physically, dissipative effects such as charge separation and mode conversion occur at these resonances, resulting in continuum damping. These effects and the resulting damping can be described using a kinetic theory of plasmas. The standard approach to calculate this damping numerically is to represent these processes using a resistive magnetohdrodynamic (\MHD ) model. In these models, continuum damping represents the limit of resistive damping as resistivity vanishes. However, this damping can also be found in non-dissipative ideal \MHD\ theory. In this case the singularity resulting from the continuum resonance must be treated in accordance with the causality condition, which implies integration around the singularity to give a negative imaginary frequency component. This is formally analagous to the treatment of Landau damping in collisionless plasmas \cite{landau_damping}.  As they have complex frequencies, the solutions to the eigenvalue problem where continuum resonance interaction occurs are more accurately described as ``quasi-modes'' \cite{Principles_of_MHD}. However, the more common description of the solutions as ``modes'' is used in this work.

A number of techniques have been developed for calculating continuum damping using ideal \MHD . Analytic treatments have been developed based on solution of the wave equation using asymptotic matching \cite{continuum_damping_of_high_n_TAWs}, representing the resonance as a perturbation following a ballooning transformation \cite{continuum_damping_of_ideal_TAEs,resonant_damping_of_TAEs_in_tokamaks} and representing the resonance as a perturbation to the Lagrangian for a global mode \cite{continuum_damping_of_low_n_TAEs}. These techniques have been limited to the analysis of \TAEs\ in large aspect-ratio circular cross-section tokamaks. Continuum damping has previously been calculated numerically by including a set of analytically determined jump conditions due to singularities at continuum resonances \cite{numerical_study_of_Alfven_continuum_damping_of_AEs}. Continuum damping has also been computed for shear Alfv\'{e}n eigenmodes in tokamak and stellarator cases using a numerical method in which the eigenvalue problem was solved along a complex contour \cite{computational_approach_to_continuum_damping_in_3D_published}. However, in these cases, calculations were resticted to only two coupled harmonics in the large aspect ratio limit. Nevertheless, this complex contour technique proved more computationally efficient than the standard approach of finding damping in the limit of vanishing resistivity.

In this paper we describe the results of a complex contour method for calculating continuum damping of shear Alfv\'{e}n eigenmodes in general three-dimensional toroidal geometry. Thus, modes involving couplings between many different poloidal and toroidal harmonics may be computed. This allows, for the first time, the capability to compute continuum damping in stellarators with realistic geometry. Section~\ref{sec:SAE} briefly describes the reduced shear Alfv\'{e}n wave equation and its solution using the code \CKA\ \cite{CKA_Thesis}. The complex contour technique used to compute the continuum damping is discussed in Section~\ref{sec:CCM}. This section outlines the choice of contour and calculation of equilibrium quantities for complex values of the radial coordinate. This technique is applied to a \TAE\ in a tokamak with circular cross-section in Section~\ref{sec:tokamak}, where it is verified against the previously developed resistive \MHD\ \cite{damping_of_GAEs_in_tokamaks_due_to_resonant_absorption} and large-aspect ratio complex contour \cite{computational_approach_to_continuum_damping_in_3D_published} continuum damping calculations. The complex contour technique is used to calculate the continuum damping of \TAEs\ and \EAEs\ in stellarators with torsatron, helias and heliac configurations in Section~\ref{sec:stellarator}.

\section{Shear Alfv{\'{e}}n eigenmodes} \label{sec:SAE}
\CKA\ (Code for Kinetic Alfv\'{e}n waves) is a finite element \MHD\ code which solves reduced \MHD\ equations for $\Phi$, the peturbation to the scalar potential \cite{CKA_Thesis}. This code can be used to find Alfv\'{e}n eigenmodes in an incompressible plasma in either two or three dimensions. The code uses Boozer coordinates $\left ( s, \theta , \phi \right )$, where $s \in \left [ 0 , 1 \right ] $ is the square root of the normalised toroidal flux $s = \sqrt{\frac{\psi}{\psi_{edge}}} $. \CKA\ represents a discretisation of the following Alfv\'{e}n wave equation using the Galerkin method
\begin{eqnarray}
\noindent -\omega^2 \left \{ \nabla \cdot \left ( \frac{1}{v_A^2} \nabla_{\perp} \Phi \left ( \mathbf{r} \right ) \right ) + \nabla_{\perp}^2 \left [ \frac{1}{v_A^2} \left ( \frac{3}{4} \rho_i^2 + \rho_s^2 \right ) \nabla_{\perp}^2 \Phi \left ( \mathbf{r} \right ) \right ] \right \} = && \nonumber \\
\noindent \nabla \cdot \left \{ \mathbf{b} \nabla_{\perp}^2 \left [ \left (  1 - \frac{\mu_0 P^{(0)}}{B^2} \right ) \mathbf{b} \cdot \nabla \Phi \left ( \mathbf{r} \right ) \right ] \right \} + \nabla \cdot \left \{ \mathbf{b} \times \mathbf{\kappa} \frac{2 \mu_0 \mathbf{b} \times \nabla P^{(0)}}{B^2} \cdot \nabla \Phi \left ( \mathbf{r} \right ) \right \} && \nonumber \\
\noindent - \nabla \cdot \left \{ \frac{\mu_0 j_{\parallel}^{(0)}}{B} \left [ \nabla \times \left ( \mathbf{b} \left ( \mathbf{b} \cdot \nabla \Phi \left ( \mathbf{r} \right ) \right ) \right ) \right ] \right \} . \label{eq:wave_equation}
\end{eqnarray}
Here $v_A = \frac{B}{\sqrt{\mu_0 n_i m_i}}$ represents the Alfv\'{e}n speed, where $n_i$ is the number density of ions and $m_i$ is their mass. Pressure is represented by $P^{(0)}$, magnetic field strength by $B$ and field line direction by the unit vector $\mathbf{b}$. The component of the current parallel to the field lines is $j_{\parallel}^{(0)}$. In this paper negligible $\beta$ is assumed, so that terms involving $P^{(0)}$ in equation (\ref{eq:wave_equation}) can be neglected. Low ion and electron temperatures are considered so that $\rho_i = \frac{\sqrt{k m_i T_i}}{q B}$ and $\rho_s =  \frac{\sqrt{k m_i T_e}}{q B}$ are much smaller than the scale length of the waves $\frac{1}{k_{\perp}} \textasciitilde R_0$. Therefore the second term on the left hand side of equation (\ref{eq:wave_equation}) which represents the parallel electric field component, can be ignored. Frequency can also be expressed as a dimensionless quantity normalised to the Alfv\'{e}n frequency, $\Omega = \frac{\omega}{v_A ^{\left ( 0 \right )} R_0} $ where $R_0$ is the mean major radius of the plasma and $v_A ^{\left ( 0 \right )}$ is the Alfv\'{e}n speed at the magnetic axis. This equation represents an eigenvalue equation of the form
\begin{equation}
L_1 \left [ \Phi \left ( s \right ) \right ] = \omega ^2 L_2 \left [ \Phi \left ( s \right ) \right ] .
\end{equation}
where $L_1$ and $L_2$ are self-adjoint differential operators for real $s$.

Solutions to the eigenvalue equation are represented using B-splines, $B_i $, to represent variation in each dimension \cite{CKA_Thesis}. B-splines are splines with support over a minimal number of knot intervals for a given order. Thus for a two-dimensional (axisymmetric) case, the scalar potential perturbations are represented as
\begin{equation}
\Phi ' \left ( s, \theta \right ) = \sum_{i=1}^{N_s} \sum_{j=1}^{N_{\theta}} c_{ij} B_i \left ( s \right ) B_j \left ( \theta \right ) ,
\end{equation}
while in the three-dimensional case, these perturbations are represented as
\begin{equation}
\Phi ' \left ( s, \theta , \phi \right ) = \sum_{i=1}^{N_s} \sum_{j=1}^{N_{\theta}} \sum_{k=1}^{N_{\phi}} c_{ijk} B_i \left ( s \right ) B_j \left ( \theta \right ) B_k \left ( \phi \right ) .
\end{equation}
These splines are defined such that knots lie on a rectangular grid \cite{CKA_Thesis}. Using the Galerkin method, the eigenvalue problem can be discretised by first substituting the above into equation (\ref{eq:wave_equation}). The resulting expression is then multiplied by $B_i \left ( s \right ) B_j \left ( \theta \right )$ (two-dimensional case) or $B_i \left ( s \right ) B_j \left ( \theta \right ) B_k \left ( \phi \right )$ (three-dimensional case) and integrated by volume. Assigning a unique index $l$ to each set of $i$, $j$ and $k$ the resulting equation can be expressed as a matrix eigenvalue problem. The numerical solution $\Phi '$ represents a projection of the exact solution $\Phi$ onto the space spanned by the basis functions.

Eigenmodes are clearly periodic in the poloidal and toroidal directions. Moreover, they are often dominated by a small number of harmonics with close poloidal and toroidal mode numbers. Therefore it can be advatageous to use a phase factor extraction method. In this case the numerical solution is expressed as $\Phi ' = \tilde{\Phi} ' e^{i \left (\tilde{m} \theta + \tilde{n} \phi \right )}$ where $\tilde{m}$ and $\tilde{n}$ are poloidal and toroidal mode numbers and $N_{fp}$ is the number of toroidal field periods. The eigenvalue problem is reformulated in terms of $\tilde{\Phi} '$, which is a linear combination of B-splines. This reduces the poloidal and toroidal resolution required for the spline basis.

\section{Complex contour method} \label{sec:CCM}
Continuum resonance damping can be calculated by solving the \TAE\ wave equation along a complex contour in the radial coordinate \cite{computational_approach_to_continuum_damping_in_3D_published}. This is analogous to the analysis of Landau damping of plasma oscillations \cite{landau_damping}. In that case dissipationless damping arising from inhomogeneity in velocity space is analysed by choosing an appropriate contour in velocity space. Poles due to continuum resonances are circumvented, and therefore singularities in the solution are avoided. Damping is represented by an imaginary component of eigenfrequency. In order to implement this technique the equilibrium quantities in equation (\ref{eq:wave_equation}) must be analytically continued to determine their values at complex values of the radial coordinate along the integration contour. The damping of \TAEs\ computed using the complex contour technique in ideal \MHD\ has previously been shown to converge to the same value as in the limit as resistivity is reduced to zero in resistive \MHD\ \cite{computational_approach_to_continuum_damping_in_3D_published}.

The causality condition specifies that poles due to continuum resonances should lie between the integration contour and the real axis. Physically, this condition ultimately arises from the requirement that any perturbation to the state of the plasma preceedes its effect on other plasma parameters. This in turn implies that modes decay rather than grow with time and hence that $\Im \left (\omega \right ) < 0$. The imaginary frequency component causes the poles to move from real to complex values of the radial coordinate.

If the integration contour is shifted away from poles due to continuum resonances, the calculated modes will vary less rapidly in their vicinity. Thus, the solution to the eigenvalue problem can be accurately represented by basis functions defined on a coarser radial grid. Consequently, the eigenvalue will have faster convergence with respect to radial grid resolution for greater deformations of the integration contour. However, excessive deformations may result in the disappearence of Alfv\'{e}n eigenmodes as the continuum gaps they exist in close. In such cases the chosen complex contour intersects with analytically continued branches of the continuous spectrum. Additionally, non-physically meaningful poles may result due to the analytic continuation of equilibrium quantities. These poles should not lie between the integration contour and real axis, in order to avoid spurious contributions to the eigenfrequency.

The continuum damping calculations described below are performed using a complex contour described by:
\begin{equation}
s = x + i f \left ( x \right ) .
\end{equation}
Here, $x \in \left [0 , 1 \right ]$ is a real parameter. The real valued function $f \left ( x \right )$ is either chosen to be a quadratic:
\begin{equation}
f \left ( x \right ) = \alpha x \left ( x - 1 \right ) \label{eq:rak_path}
\end{equation}
or the product of a quadratic and a Gaussian function
\begin{equation}
f \left ( x \right ) = \alpha x \left ( x - 1 \right ) e^{-\left (\frac{ x - x_\gamma}{x_\beta} \right )^2} . \label{eq:awk_path}
\end{equation}
The parameters $\alpha$, $x_\beta$ and $x_\gamma$ respectively determine the scale, width and location of the deformation into the complex plane. Defining a profile using the latter function allows a more localised contour deformation than the former. Thus, larger deformations in the vicinity of the pole can occur while avoiding non-physically meaningful poles. Therefore, it is possible to calculate continuum damping with a smaller grid resolution. However, as the deformation becomes narrower and its scale increases, the rate of variation of the solution in that region is expected to increase. Thus, increased radial resolution may be required where $\alpha$ is too large and $x_\beta$ is too small.

\subsection{Equilibrium quantity representation}
In order to solve the eigenvalue problem in equation (\ref{eq:wave_equation}) it is necessary to calculate values of equilibrium quantities for complex values of $s$. These include density $\rho$, toroidal flux $\psi$, poloidal flux, $\psi_p$, current $\mathbf{j}$, magnetic field, $\mathbf{B}$, pressure $P^{\left ( 0 \right )}$, Jacobian $\sqrt{g}$ and metric tensor elements $g^{ij}$. One possible approach to estimating these quantities for complex $s$ is by using a truncated Taylor series expansion. Where the radial coordinate $s$ has real component $s_r$ and imaginary component $s_i$, expanding the function $y\left ( s \right )$ about $s_r$ yields
\begin{equation}
y(s) = y(s_r) + if\left ( s_r \right )y'\left ( s_r \right ) - \frac{1}{2}\left (f \left ( s_r \right ) \right )^2 y''\left ( s_r \right ) - \frac{i}{6}\left (f \left ( s_r \right ) \right )^3 y'''\left ( s_r \right )\cdots .
\end{equation}
This approximation may be used with spline representations of equilibrium quantities. However, these piecewise polynomial functions are not analytic, which can lead to branch cuts when extended to the complex plane. If the complex contour deviates sufficiently from the real axis, the calculated values of the functions might contain significant contour dependent errors. Integration across branch cuts may thus affect the result of the finite element calculation, leading to poor eigenvalue convergence with respect to the chosen complex contour. This is found to be the case for both the axisymmetric tokamak and stellarator examples, when the first three terms in the expansion for each quantity were retained. In these cases poor convergence is observed with respect to the choice of contour (as parameterised $\alpha$, $x_\beta$ and $x_\gamma$ for contours described by equation (\ref{eq:awk_path})).

Consequently, in these calculations real and imaginary components of equilibrium quantities along the chosen complex contour are each represented in terms of cubic B-splines. These piecewise representations are determined from analytic representations of the equilibium quantities. Equilibrium data from the \VMEC\ code \cite{steepest_descent_moment_method} are used to obtain a Fourier series in $\theta$ and $\phi$ of polynomial functions in terms of $s$. These polynomial functions are calculated based on the equilibrium data using a least squares fitting routine, ignoring unreliable data close to the magnetic axis. The use of polynomial functions to represent equilibrium quantities allows trivial analytic continuation from the real axis to determine their values at complex $s$. The aforementioned spline representations of the non-flux function equilibrium quantities are calculated by evaluating these series at positions along the chosen complex contour in $s$ and at various values of $\theta$ and $\phi$. Use of spline representations for these quantities is advantageous as the evaluation of polynomials is relatively computationally inexpensive.

\section{Tokamak results} \label{sec:tokamak}
\begin{figure}[h] 
\centering 
\includegraphics[width=80mm]{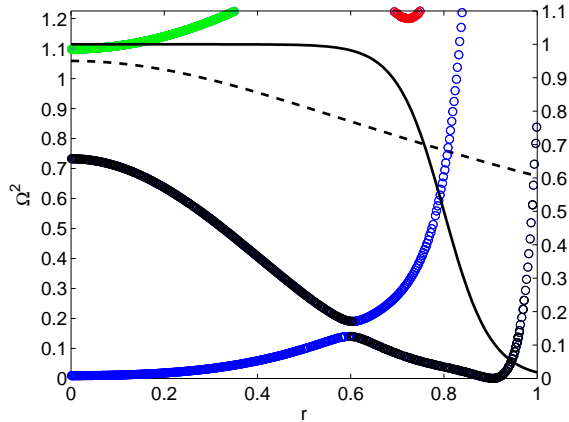}
\caption{\label{fig:tok_continuum} 
Shear Alfv\'{e}n continuous spectrum as a function of radial position for a circular-cross section tokamak computed using \CKA . The avoided crossing between the $\left ( m = 2 , n = -2 \right )$ branch ($\textcolor{RoyalBlue}{\circ}$) and the $\left ( 3 , -2 \right )$ branch ($\circ$) result in a \TAE\ (not shown). The rotational transform profile, $\iota$, (dashed line) and normalised density profile, $\frac{n_i}{n_{i0}}$, (solid line) are also plotted with the scale shown on the right axis.
}
\end{figure}

The complex contour technique for calculating continuum damping was applied to a \TAE\ in an axisymmetric tokamak with circular cross-section. An inverse aspect ratio of $\epsilon = 0.225$ is chosen for this case. The rotational transform is described by the polynomial profile $\iota \left (s \right ) = 0.95016 - 0.67944 s^2 + 0.62286 s^4 - 0.41244 s^6 + 0.1219 s^8 + 0.0042185 s^{10} -0.0013979 s^{12}$. Here the rotational transform is defined as $\iota = \lim_{\Delta \phi \to \infty} \frac{\Delta \theta}{\Delta \phi}$ where $\Delta \theta$ and $\Delta \phi$ are respectively the displacement in the poloidal and toroidal direction along a magnetic field line. The density profile is chosen to be
\begin{equation}
n_i \left (s \right )=\frac{n_{i0}}{2} \left (1-\tanh \left (\frac{s-\Delta_1}{\Delta_2}\right ) \right ), \label{eq:density}
\end{equation}
where $\Delta_1 = 0.8$ and $\Delta_2 = 0.1$. This approximates the choice of $\iota$ and $n_i$ used by K\"{o}nies and Kleiber \cite{computational_approach_to_continuum_damping_in_3D_published}. The functions $n_i \left (s \right ) $ and $\iota \left (s \right ) $ are overlaid on the resulting continuous spectrum shown in figure~\ref{fig:tok_continuum}. This equilibrium results in a \TAE\ due primarily to the coupling of the $\left ( m , n \right ) = \left ( 2 , -2 \right )$ and $\left ( 3 , -2 \right )$ harmonics. Complex contours of the form indicated in equation (\ref{eq:rak_path}) were used.

\begin{figure}[h] 
\centering 
\includegraphics[width=80mm]{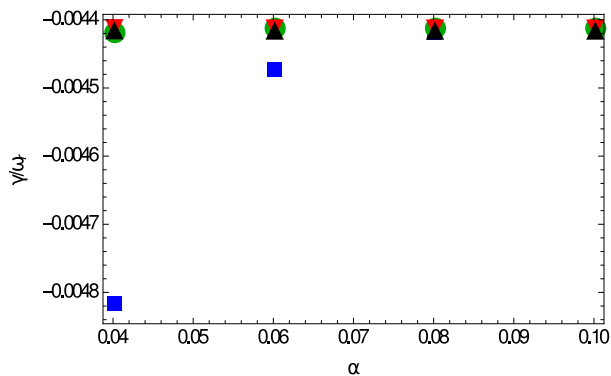}
\caption{\label{fig:tok_cc_alpha_con} 
Convergence of continuum damping with respect to the contour deformation parameter $\alpha$. Damping ratio is plotted as a function of $\alpha$ for $N_s = 400$ and $N_{\theta} = 20$ ($\textcolor{RoyalBlue}{\blacksquare}$), $N_s = 800$ and $N_{\theta} = 20$ ($\textcolor{OliveGreen}{\bullet}$), $N_s = 1200$ and $N_{\theta} = 20$ ($\textcolor{Red}{\blacktriangledown}$) and $N_s = 1200$ and $N_{\theta} = 10$ ($\textcolor{Black}{\blacktriangle}$).
}
\end{figure}

The frequency of this \TAE\ was calculated for various values of the contour deformation parameter $\alpha$ and radial grid resolution $N_s$. For this axisymmetric plasma the solution is represented by splines defined on a two-dimensional rectangular grid in $s$ and $\theta$. The results of this calculation are plotted in figure~\ref{fig:tok_cc_alpha_con}, demonstrating convergence of continuum damping with respect to these parameters. Thus, a complex normalised \TAE\ frequency of $\Omega = 0.39515 -0.00174465 i$ is obtained, resulting in a damping ratio of $\frac{\gamma}{\omega_r} = -0.00441515$. This convergence indicates that sufficiently large values of $\alpha$ and $N$ are used that the mode structure is resolved. The most demanding element of this structure in terms of radial grid resolution is the continuum resonance interaction, particularly where the chosen contour passes close to the pole due to the continuum resonance. In that region there is rapid variation in the perturbation to the scalar potential due to the mode, requiring a large density of grid points to resolve. A reduction in the poloidal grid resolution from $N_{\theta} = 20$ to $10$ results in a change in damping ratio of less than $0.07 \% $, indicating satisfactory convergence with respect to this parameter also.

\begin{figure}[h] 
\centering 
\includegraphics[width=80mm]{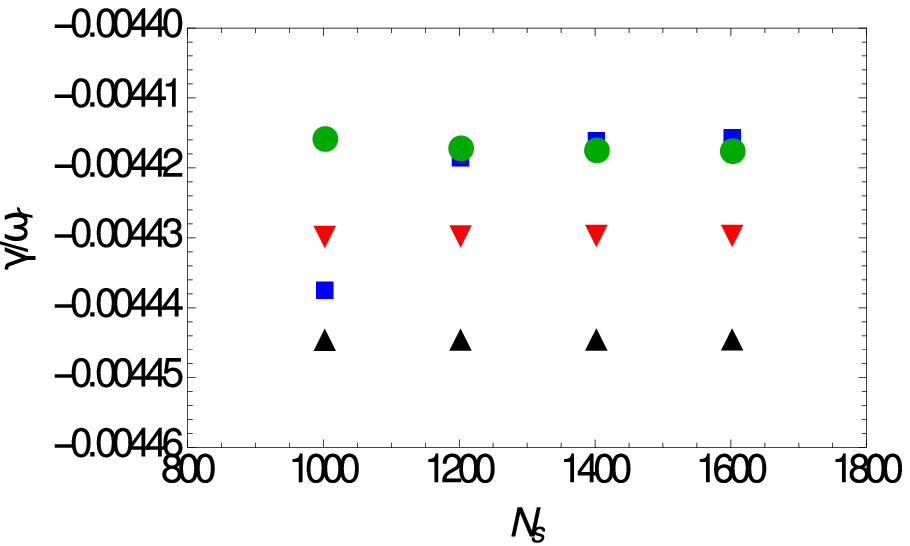}
\caption{\label{fig:tok_res_rad_con} 
Convergence of damping as the artificial resistivity parameter $\delta$ is reduced for a circular cross-section tokamak. Damping ratio is plotted as a function of $N_s$ for $\delta = 5 \times 10^{-9}$ ($\textcolor{RoyalBlue}{\blacksquare}$), $1 \times 10^{-8}$ ($\textcolor{OliveGreen}{\bullet}$), $5 \times 10^{-8}$ ($\textcolor{Red}{\blacktriangledown}$) and $1 \times 10^{-7}$ ($\textcolor{Black}{\blacktriangle}$).
}
\end{figure}

Continuum damping of Alfv\'{e}n eigenmodes in axisymmetric tokamaks has been computed previously by determining the limit of resistive damping as resistivity tends to zero \cite{damping_of_GAEs_in_tokamaks_due_to_resonant_absorption}. Therefore, the resistive calculation can be used as a benchmark for the complex contour calculation. This calculation is implemented in \CKA\ by inserting an additional term on the left hand side of the equation (\ref{eq:wave_equation}),
\begin{equation}
-i \delta \nabla_\perp^2 \left ( \frac{\omega^2}{v_A^2} \nabla_\perp^2 \left ( \Phi \left ( s \right ) \right ) \right )
\end{equation}
Damping is plotted as a function of radial grid resolution for a series of different values of the artificial damping parameter $\delta$ in figure~\ref{fig:tok_res_rad_con}. This plot indicates convergence in the damping ratio with respect to both $\delta$ and $N$. Complex normalised \TAE\ frequency is estimated to be $\Omega = 0.395151 - 0.00174521 i$ corresponding to a damping ratio of $\frac{\gamma}{\omega_r} = -0.00441658$. In this case the density of the grid in the radial coordinate is increased in the vicinity of the continuum resonance at $s = 0.964$, to increase the rate of convergence. Satisfactory convergence can also be demonstrated with respect to the poloidal coordinate, as the change in the damping ratio computed when the poloidal grid resolution is reduced from $N_{\theta} = 20$ to $10$ is less than $0.07 \%$.

The small difference between the damping calculated using the complex contour and artificial resistivity methods can be attributed to the errors present in each. Approximation of equilibrium quantities in terms of cubic splines may account for the variation in the continuum damping ratio with contour deformation $\alpha$. In the artificial resistivity method a finite resistivity is required to resolve the resonant interaction with the continuum due to the finite radial grid resolution \cite{global_waves_in_cold_plasma}. This results in a small resistive damping occuring throughout the plasma in addition to the continuum damping very close to the magnetic surface on which the continuum resonance occurs. Reducing the resistivity to improve accuracy requires a greater radial grid resolution to obtain convergence, as the finite width of the peak in the mode due to continuum resonance must be adaquately resolved. For sufficiently large $\alpha$ the damping found using the complex contour technique is found to converge more rapidly with respect to $N_s$ than for the resistive technique.

It is also interesting to compare the continuum damping calculated in the above fully two-dimensional case to that found using the large aspect-ratio approximation. This approximation results in a differential equation derived by Berk \textit{et al.} to describe \TAEs\ \cite{continuum_damping_of_low_n_TAEs}. The application of the complex contour technique based on this formula has been described previously by K\"{o}nies and Kleiber \cite{computational_approach_to_continuum_damping_in_3D_published} and Bowden \textit{et al.} \cite{comparison_of_methods_for_numerical_calculation_of_continuum_damping}. Using the shooting method described in the latter and considering only the $\left ( 2 , -2 \right )$ and $\left ( 3 , -2 \right )$ harmonics, a \TAE\ is found with normalised frequency $\Omega = 0.370887 - 0.00174204 i$ corresponding to a damping ratio of $\frac{\gamma}{\omega_r} = -0.00469695$. Thus the calculation based on the large aspect ratio approximation agrees reasonably closely with the fully two dimensional calculation in this case (to within $6.6 \%$). This suggests that modifications to \TAEs\ due to terms of higher order with respect to aspect ratio and couplings to additional harmonics do not have a large effect on their continuum damping in this case.

The three-dimensional continuum damping calculation was verified through application of the fully three-dimensional calculation to this tokamak case. That is, the quadratic form was computed using integrals for basis splines defined on a three-dimensional grid. In the resulting solution, toroidal variation of eigenmodes is represented by splines with a radial grid resolution of $N_{\phi} = 10$ in this dimension and $N_s = 400$ and $N_{\theta} = 10$. This results in a \TAE\ frequency which agrees with the corresponding two-dimensional calculation to $7$ significant figures, the precision provided in the output of the code.

\section{Stellarator results} \label{sec:stellarator}

\subsection{Torsatron eigenmodes}
\begin{figure}[h] 
\centering 
\includegraphics[width=80mm]{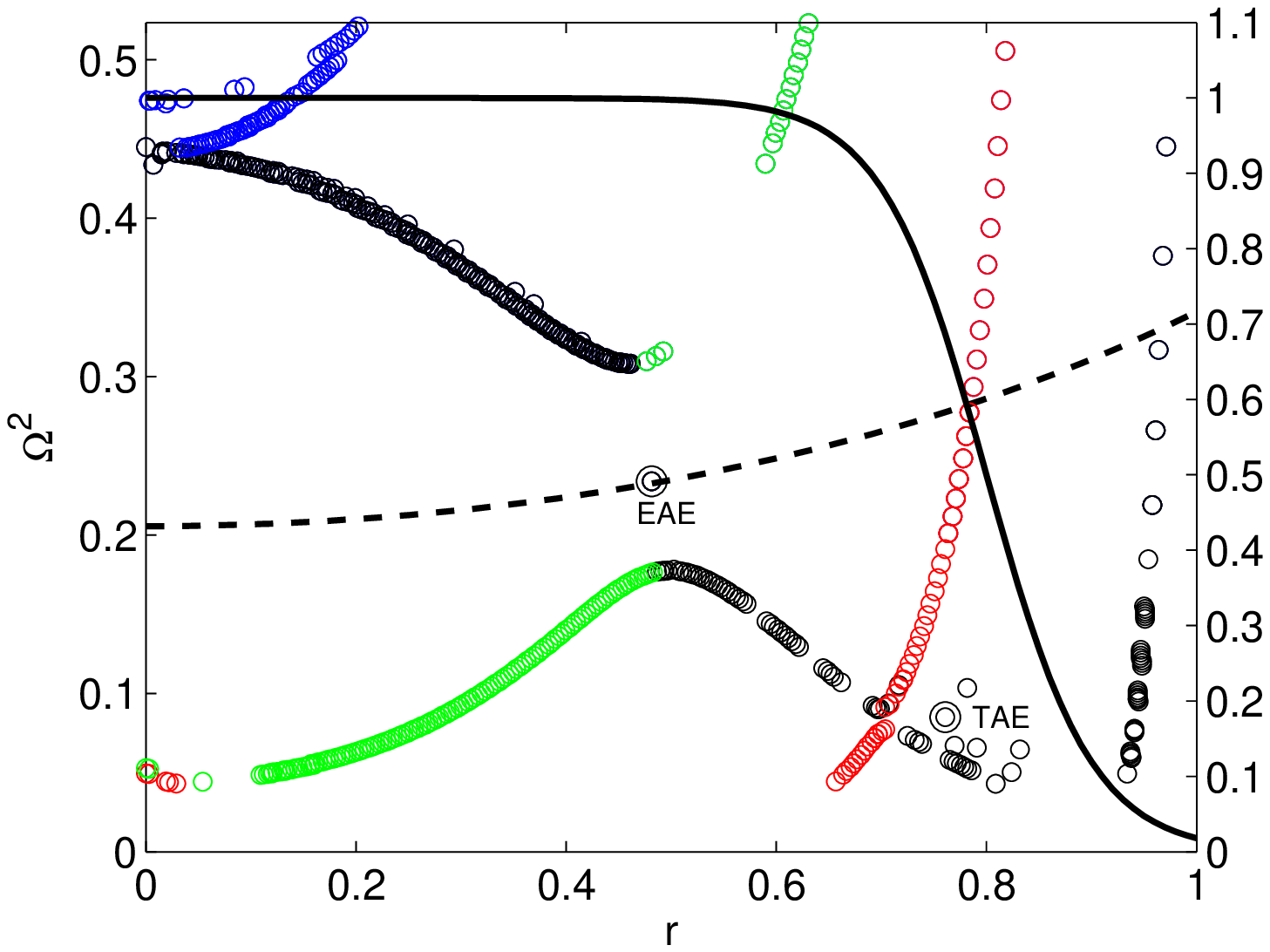}
\caption{\label{fig:torsatron_continuum} 
Shear Alfv\'{e}n continuous spectrum as a function of radial position for the torsatron case computed using \CKA . The avoided crossing between the $\left ( m = 3 , n = -2 \right )$ branch ($\circ$) and the $\left ( 4 , -2 \right )$ branch ($\textcolor{Red}{\circ}$) produce a \TAE\ in the resulting spectral gap. Likewise the avoided crossing between the former branch and the $\left ( 5 , -2 \right )$ branch ($\textcolor{Green}{\circ}$) gives rise to an \EAE\ in the corresponding spectral gap. The locations of the maxima of these global modes are shown, approximately corresponding to the radial locations of the gaps. The rotational transform profile, $\iota$, (dashed line) and normalised density profile, $\frac{n_i}{n_{i0}}$, (solid line) are plotted with the scale shown on the right axis.
}
\end{figure}

The continuum damping of a \TAE\ and an \EAE\ in a torsatron configuration were calculated. The rotational transform for this case is defined by the polynomial $\iota \left (s \right ) = 0.4319 + 0.23407 s^2 + 0.042125 s^4 + 0.008341 s^6 $. This profile is typical of torsatron configurations, with a minimum near the magnetic axis and strong magnetic shear near the edge \cite{energetic_ion_driven_global_instabilities_in_stellarators}. For simplicity, the same density profile is used here as in the tokamak case in Section~\ref{sec:tokamak}. The stellarator has $N_{fp} = 20$ toroidal field periods with a major radius of $R = 20 \textup{m}$. The plasma cross section is elliptical, with semi-major axis of $a = 1.13 \textup{m}$ and semi-minor axis of $b = 0.867 \textup{m}$. \TAEs\ have previously been observed in torsatrons such as the Large Helical Device (LHD ) \cite{energetic_ion_driven_MHD_instabilities_in_CHS_and_LHD} and the Compact Helical System (\CHS ) \cite{energetic_ion_driven_TAE_in_a_heliotron,energetic_ion_driven_MHD_instabilities_in_CHS_and_LHD}. In some instances continuum damping has been identified as the dominant cause of damping for these modes \cite{excitation_of_TAEs_in_a_heliotron}. In the present case a large \EAE\ gap exists due to the elongation of the plasma, while the \TAE\ gap is small due to its large aspect ratio. The continuous spectrum is plotted as a function of radial position in figure~\ref{fig:torsatron_continuum}, demonstrating these features.

A \TAE\ was found due to coupling between the $\left ( 3 , -2 \right )$ and $\left ( 4 , -2 \right )$ harmonics. This \TAE\ has resonances with $n = 2$ branches of the continuum at $r \approx 0.28$ and $r \approx 0.94$. Using complex contours described by equation (\ref{eq:awk_path}) it is possible to calculate the contribution of each resonance to the overall damping separately by localising the contour deformation near that resonance. This approach assumes that the other continuum resonance interaction does not significantly alter the mode structure. This is not always the case, even for low values of damping \cite{comparison_of_methods_for_numerical_calculation_of_continuum_damping}. However, in this example we will assume that damping due to the resonance near the edge is much larger than that near the core. This is expected, as the avoided crossing of the $\left ( 3 , -2 \right )$ and $\left ( 4 , -2 \right )$ continuum branches is closer to the resonance with the $\left ( 4 , -2 \right )$ branch, so the mode would be expected to have greater amplitude at this crossing. Moreover, the amplitude of the $\left ( 4 , -2 \right )$ harmonic is expected to be far larger than the $\left ( 5 , -2 \right )$ harmonic throughout the plasma. Thus, $x_{\gamma} = 0.96$ is initially chosen, localising the contour deformation near the resonance close to the edge of the plasma.

\begin{figure}[h] 
\centering 
\includegraphics[width=80mm]{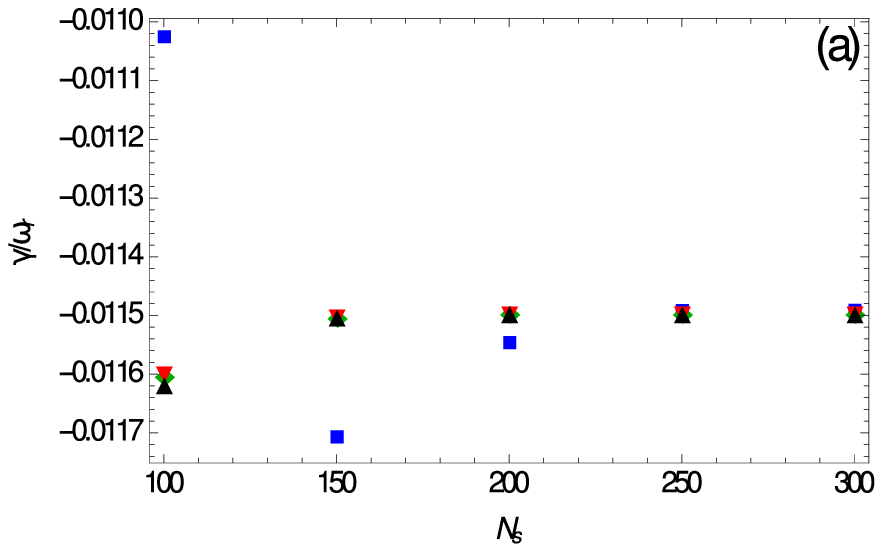}
\includegraphics[width=80mm]{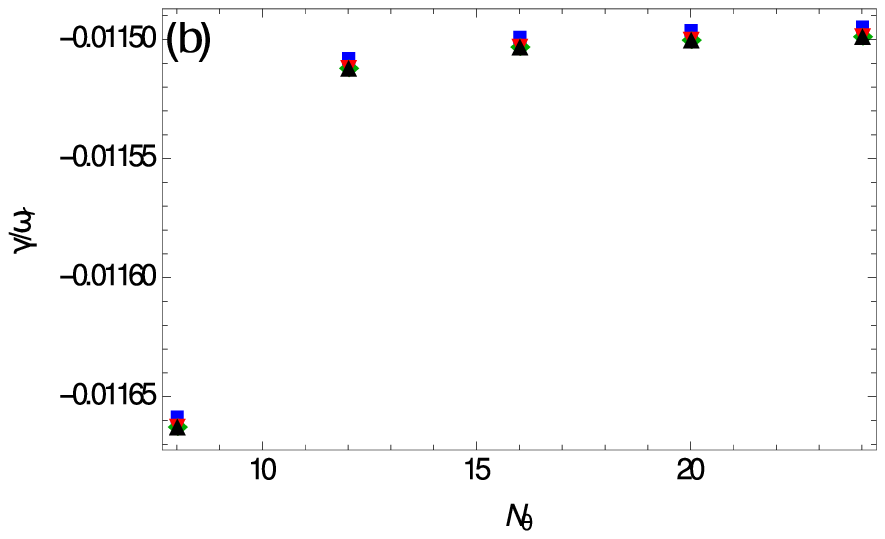}
\includegraphics[width=80mm]{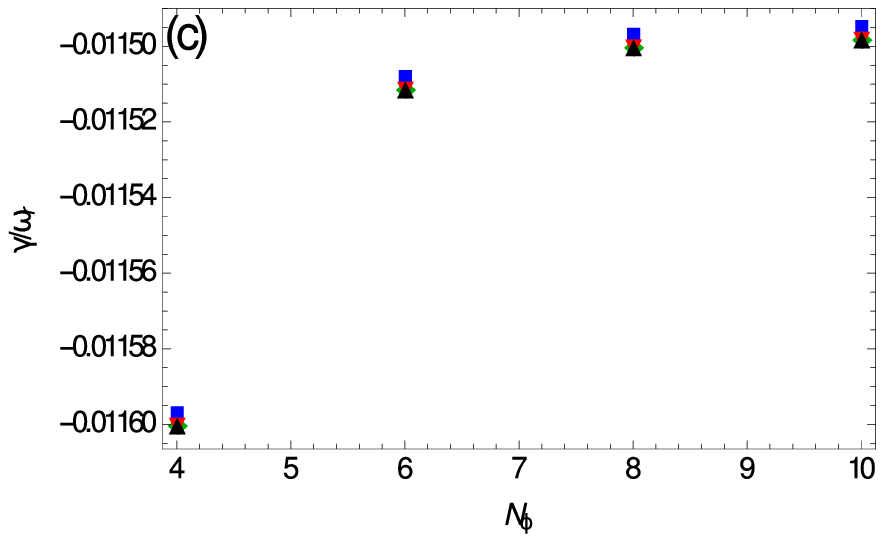}
\caption{\label{fig:tor_TAE_con} 
Convergence of continuum damping ratio for a \TAE\ due to coupling of the $\left ( 3 , -2 \right )$ and $\left ( 4 , -2 \right )$ shear Alfv\'{e}n wave harmonics in a torsatron plasma, with respect to (a) radial grid resolution $N_s$, (b) poloidal grid resolution $N_{\theta}$ and (c) toroidal grid resolution $N_{\phi}$. In each case, two of $N_s = 300$, $N_{\theta} = 20$ and $N_{\phi} = 8$ are fixed and the other parameter is varied. Complex contours are chosen with parameters $x_\beta = 0.1$ and $x_\gamma = 0.96$. Damping ratio is plotted for $\alpha = 0.1$ ($\textcolor{RoyalBlue}{\blacksquare}$), $\alpha = 0.2$ ($\textcolor{OliveGreen}{\bullet}$), $\alpha = 0.5$ ($\textcolor{Red}{\blacktriangledown}$) and $\alpha = 1.0$ ($\textcolor{Black}{\blacktriangle}$).
}
\end{figure}

The damping ratio of the \TAE\ is plotted as a function of $N_s$, $N_{\theta}$ and $N_{\phi}$ for various values of $\alpha$, $x_{\beta} = 0.1$ and $x_{\gamma} = 0.96$ in figure~\ref{fig:tor_TAE_con}. The ratio $\frac{\gamma}{\omega_r}$ is found to be approximately constant with respect to each of these parameters, indicating convergence with respect to grid resolution in each dimension and the complex contour chosen. The most converged estimate of the complex normalised \TAE\ frequency obtained is $\Omega = 0.29031 - 0.00333875 i $ corresponding to a damping ratio of $\frac{\gamma}{\omega_r} = -0.0115006$. When the complex deformation is localised near the inner resonance and grid resolution is kept constant, comparatively weak damping is found. For $\alpha = 0.1$ to $0.5$, $x_\beta = 0.05$ and $x_\gamma = 0.28$ a complex normalised frequency of $\Omega \approx 0.291398 - 2.7 \times 10^{-8} i $ and a damping ratio of $\frac{\gamma}{\omega_r} \approx -9.2 \times 10^{-8}$ are computed. This validates the assumption that the continuum interaction with the $\left ( 5 , -2 \right )$ branch is not significant for this mode.

\begin{figure}[h] 
\centering 
\includegraphics[width=80mm]{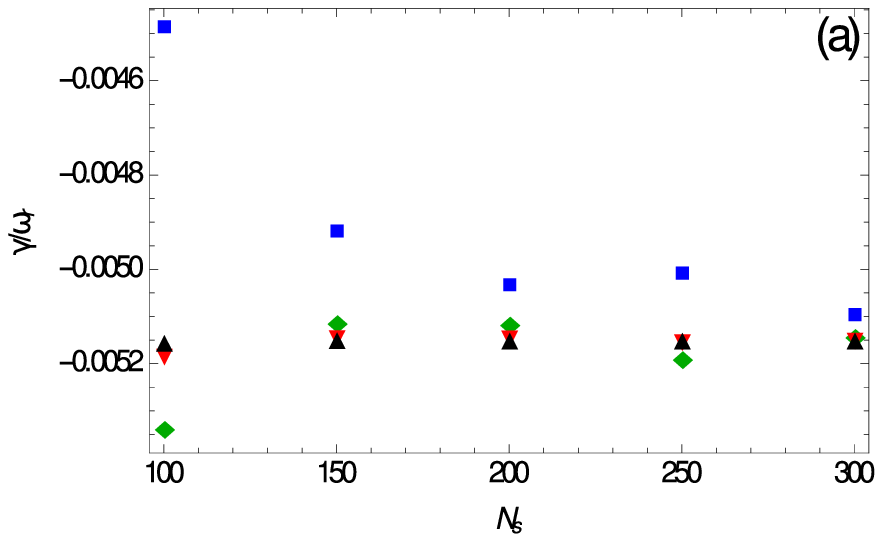}
\includegraphics[width=80mm]{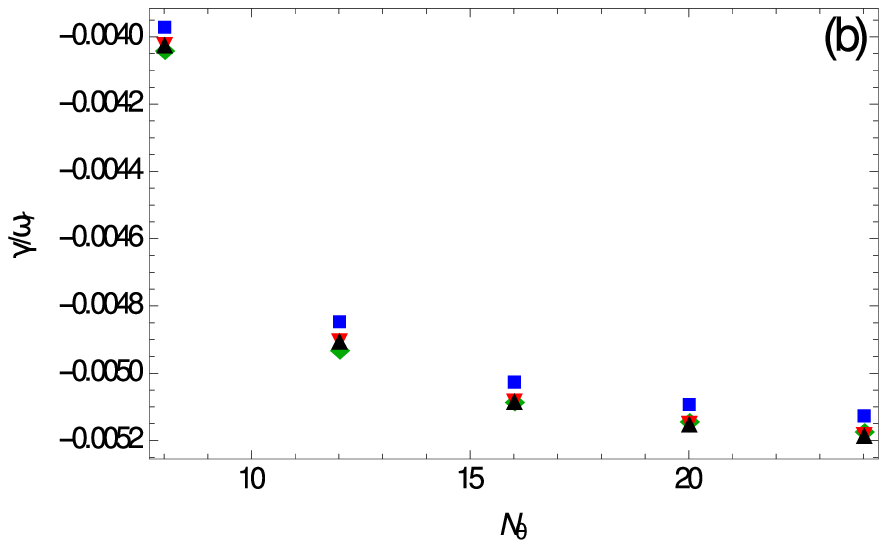}
\includegraphics[width=80mm]{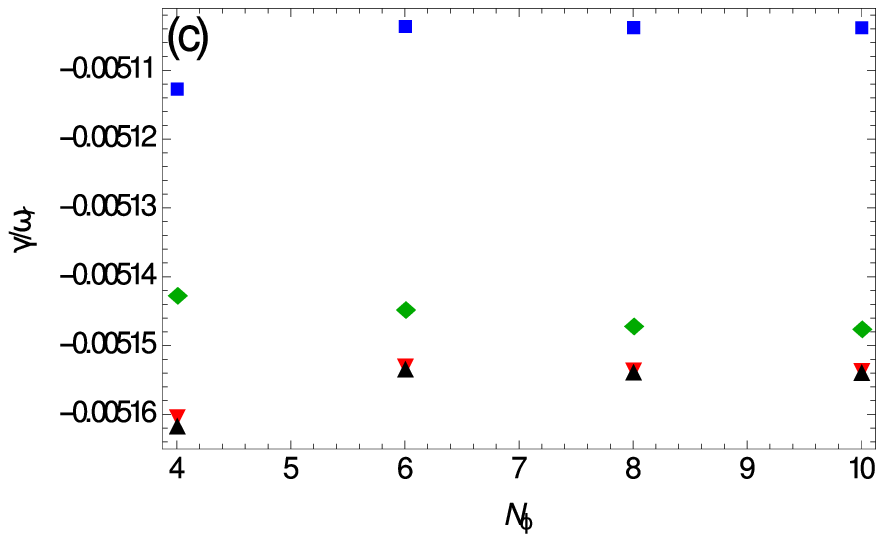}
\caption{\label{fig:tor_EAE_con} 
Convergence of continuum damping ratio for an \EAE\ due to coupling of the $\left ( 3 , -2 \right )$ and $\left ( 5 , -2 \right )$ shear Alfv\'{e}n wave harmonics in a torsatron plasma, with respect to (a) radial grid resolution $N_s$, (b) poloidal grid resolution $N_{\theta}$ and (c) toroidal grid resolution $N_{\phi}$. In each case, two of $N_s = 300$, $N_{\theta} = 20$ and $N_{\phi} = 8$ are fixed and the other parameter is varied. Complex contours are chosen with parameters $x_\beta = 0.1$ and $x_\gamma = 0.96$. Damping ratio is plotted for $\alpha = 0.1$ ($\textcolor{RoyalBlue}{\blacksquare}$), $\alpha = 0.2$ ($\textcolor{OliveGreen}{\bullet}$), $\alpha = 0.5$ ($\textcolor{Red}{\blacktriangledown}$) and $\alpha = 1.0$ ($\textcolor{Black}{\blacktriangle}$).
}
\end{figure}

The \EAE\ for which continuum damping is calculated is due to coupling of the $\left ( 3 , -2 \right )$ and $\left ( 5 , -2 \right )$ harmonics. This mode has resonances with the $\left ( 3 , -2 \right )$ continuum branch at $s \approx 0.78 $ and with the $\left ( 4 , -2 \right )$ continuum branch at $s \approx 0.96 $. Once again, the contribution of each continuum resonance to the damping can be computed separately. As expected, it is interaction with the latter branch that contributes most to continuum damping. The convergence of the damping ratio due to the outer resonance, obtained using $x_{\gamma} = 0.96$ is plotted in figure~\ref{fig:tor_EAE_con}. The converged case corresponds to a complex normalised frequency of $\Omega = 0.48197 - 0.00248407 i$ and hence a damping ratio of $\frac{\gamma}{\omega_r} = -0.00515400$ for the outer resonance. In contrast, examining the effect of the inner continuum resonance by setting $x_{\gamma} = 0.78$ and varying between $\alpha = 0.1$ and $\alpha = 0.5$ results in a complex normalised frequency of $\Omega = 0.484171 - 0.0000149544 i$ and a damping ratio of just $\frac{\gamma}{\omega_r} \approx -0.000031$. Thus, the contribution of the inner continuum resonance to the damping of the mode is very small compared with the outer one and can be safely neglected.

\begin{figure}[h] 
\centering 
\includegraphics[width=80mm]{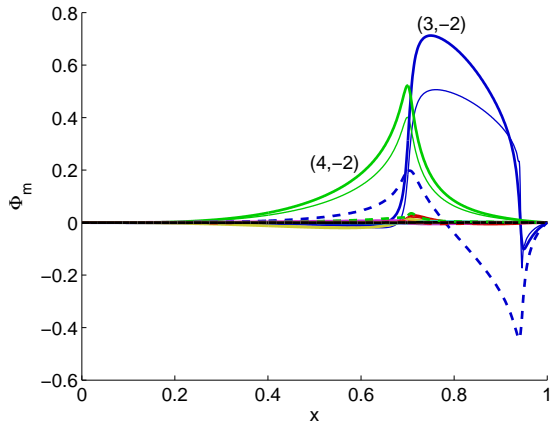}
\caption{\label{fig:torsatron_TAE}
Largest Fourier harmonics for the \TAE\ identified in the torsatron case as a function of contour parameter $x$. The dominant $\left ( 3 , -2 \right )$ (blue) and $\left ( 4 , -2 \right )$ (green) harmonics are labelled. The mode calculated along a complex contour of the form indicated in equation~\ref{eq:awk_path} with $x_{\alpha} = 0.1$, $x_{\beta} = 0.1$ and $x_{\gamma} = 0.96$ (thick line) is compared with that calculated for real $s$ (thin line). The real component is indicated by the solid line and the imaginary component is indicated by the dashed line.
}
\end{figure}

\begin{figure}[h] 
\centering 
\includegraphics[width=80mm]{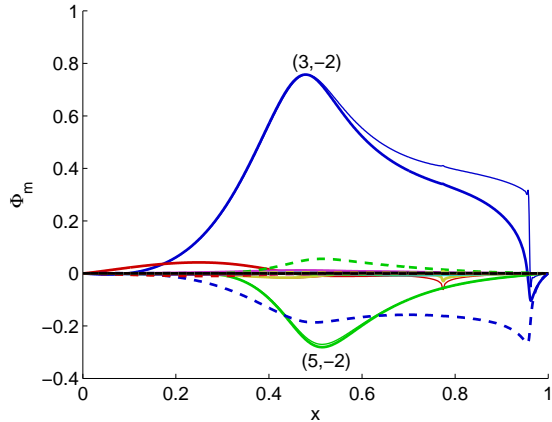}
\caption{\label{fig:torsatron_EAE}
Largest Fourier harmonics for the \EAE\ identified in the torsatron case as a function of contour parameter $x$. The dominant $\left ( 3 , -2 \right )$ (blue) and $\left ( 5 , -2 \right )$ (green) harmonics are labelled. The mode calculated along a complex contour of the form indicated in equation~\ref{eq:awk_path} with $x_{\alpha} = 0.1$, $x_{\beta} = 0.1$ and $x_{\gamma} = 0.96$ (thick line) is compared with that calculated for real $s$ (thin line). The real component is indicated by the solid line and the imaginary component is indicated by the dashed line.
}
\end{figure}

The rapid convergence with respect to $N_{\theta}$ and $N_{\phi}$ in both the \TAE\ and \EAE\ cases indicates that there is relatively little coupling to poloidal and toroidal harmonics beyond the main two indicated indicated for those modes. This is confirmed based on the mode structures obtained in each case. The largest harmonics in the Fourier decomposition of the \TAE\ and \EAE\ are shown in figure~\ref{fig:torsatron_TAE} and figure~\ref{fig:torsatron_EAE} respectively. It is found that the mode structure obtained on the complex contour is similar to that obtained solving the problem for real $s$. The largest differences in the amplitude and phase of the modes calculated in each case are observed for the $\left ( 3 , -2 \right )$ harmonic. These differences can be seen to extend well beyond the region where the contour deformation is localised, and are thought to be an effect of the continuum resonances for that harmonic.

\subsection{Helias eigenmode}
Similarly, a \TAE\ was identified and its continuum damping calculated for a \WSX\ stellarator configuration. This device is an $N_{fp} = 5$ toroidal field period helias which has a relatively complicated magnetic geometry produced by modular coils. Consequently, many different Fourier components contribute significantly to equilibrium quantities. These cause significant couplings between numerous different harmonics. However, as the rotational transform $\iota$ remains close to unity throughout the plasma, only avoided crossings due to continuum branches with relatively high $\left ( m , n \right )$ occur in \WSX . While these modes involve larger poloidal and toroidal harmonics, the relatively low magnetic shear allows such modes to have a significant radial extent.

\begin{figure}[h] 
\centering 
\includegraphics[width=80mm]{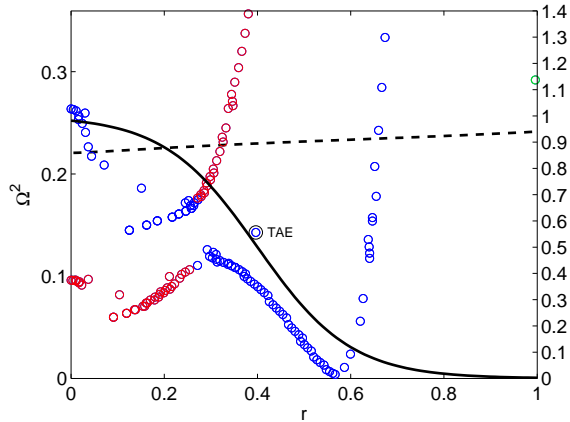}
\caption{\label{fig:W7X_Continua} 
Shear Alfv\'{e}n continuous spectrum as a function of radial position for the \WSX\ case computed using \CKA . The avoided crossing between the $\left ( m = 18 , n = -16 \right )$ branch ($\textcolor{RoyalBlue}{\circ}$) and the $\left ( 19 , -16 \right )$ branch ($\textcolor{Red}{\circ}$) leads to a \TAE\ in the resulting spectral gap. The locations of the maxima of these global modes are shown, approximately corresponding to the radial locations of the gaps.
}
\end{figure}

\begin{figure}[h] 
\centering 
\includegraphics[width=80mm]{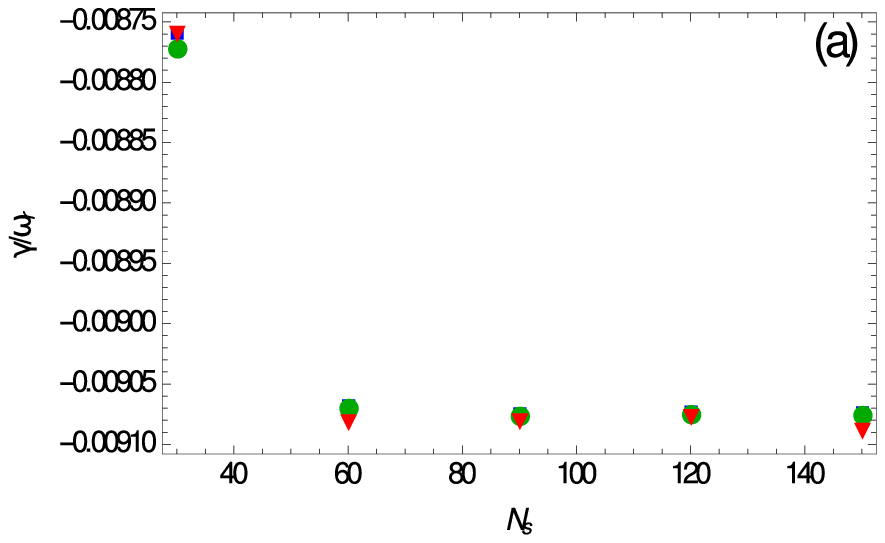}
\includegraphics[width=80mm]{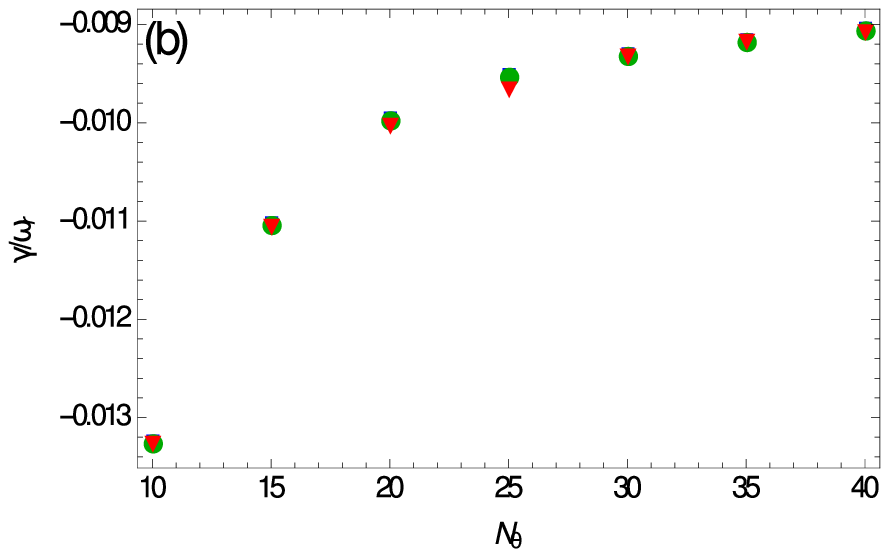}
\includegraphics[width=80mm]{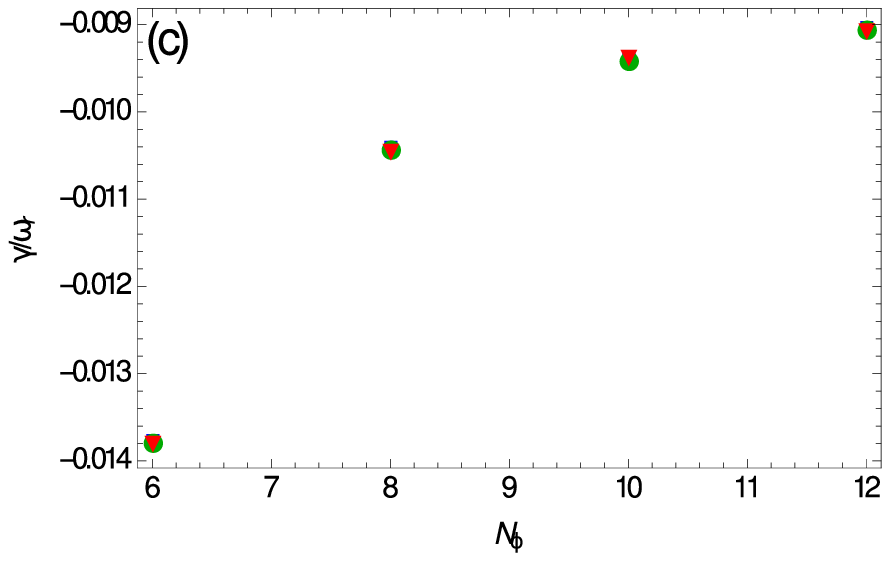}
\caption{\label{fig:W7X_TAE_con}
Convergence of continuum damping ratio for a \TAE\ due to coupling of the $\left ( 18 , -16 \right )$ and $\left ( 19 , -16 \right )$ shear Alfv\'{e}n wave harmonics in a \WSX\ plasma, with respect to (a) radial grid resolution $N_s$, (b) poloidal grid resolution $N_{\theta}$ and (c) toroidal grid resolution $N_{\phi}$. In each case, two of $N_s = 150$, $N_{\theta} = 40$ and $N_{\phi} = 12$ are fixed and the other parameter is varied. Complex contours are chosen with parameters $x_\beta = 0.1$ and $x_\gamma = 0.65$. Damping ratio is plotted for $\alpha = 0.2$ ($\textcolor{RoyalBlue}{\blacksquare}$), $\alpha = 0.5$ ($\textcolor{OliveGreen}{\bullet}$) and $\alpha = 1.0$ ($\textcolor{Red}{\blacktriangledown}$).
}
\end{figure}

A \TAE\ is found which is primarily due to coupled $\left ( 18 , -16 \right )$ and $\left ( 19 , -16 \right )$ harmonics. The global mode calculation is done for a density profile defined using the equation in the Section~\ref{sec:tokamak}, with parameters changed to $\Delta_1 = 0.4$ and $\Delta_2 = 0.2$. This density profile is expected to reduce the effect of poloidal mode couplings compared with that used in the tokamak and torsatron cases, as the continuum resonance is moved inward. Thus the required poloidal resolution is reduced, making the calculation more tractable. By reducing $\Delta_1 $ the resonance is shifted inward to where the \TAE\ amplitude is greater and by increasing $\Delta_2 $ the radial variation in continuum frequency is decreased. A \TAE\ mode and the corresponding branches of the continuous spectrum are shown in figure~\ref{fig:W7X_Continua}. The \TAE\ has a continuum resonance with the $\left ( 18 , -16 \right )$ branch at $s \approx 0.67$. Although other continuum resonances are expected near the edge of the plasma due to the upward shift in continuum frequencies resulting from decreasing plasma density, it is believed that this resonance represents the dominant contribution to the continuum damping of this mode. This assumption is supported by the continuum damping calculations for the torsatron, where it is shown that continuum damping is dominated by interaction with branches corresponding to the dominant harmonics of the global mode. Other \TAEs\ with the same dominant harmonics and a different number of nodes are found in the gap.

The computed damping ratio for the selected \TAE\ is plotted as a function of $N_s$, $N_{\theta}$ and $N_{\phi}$ in figure~\ref{fig:W7X_TAE_con}. These plots show convergence with respect to the number of splines used in each dimension, as well as the contour deformation parameter $\alpha$. Thus the normalised complex frequency is estimated to be $\Omega = 0.394342 - 0.00047676 i $ based on the value obtained for the highest grid resolution in each dimension and smallest contour deformation. This corresponds to a continuum damping ratio of $\frac{\gamma}{\omega_r} = -0.0012090001 $. Fine poloidal and toroidal resolution is required in this case due to strong couplings between the $\left ( m , n \right )$ and $\left ( m' , n' \right )$ harmonics for a range of different $m' - m$ and $n' - n$. This is in contrast to the torsatron case where couplings for $m' - m = \pm 1$ and $n' - n = 0$ harmonics (toroidicity induced coupling) and $m' - m = \pm 2$ and $n' - n = 0$ harmonics (elipticity induced coupling) are expected to be most significant and damping can be computed relatively accurately for lower poloidal and toroidal resolution. Thus we conclude that the continuum damping in the \WSX\ case is significantly affected by coupling to a range of non-dominant harmonics.

\subsection{Heliac eigenmode}
\begin{figure}[h] 
\centering 
\includegraphics[width=80mm]{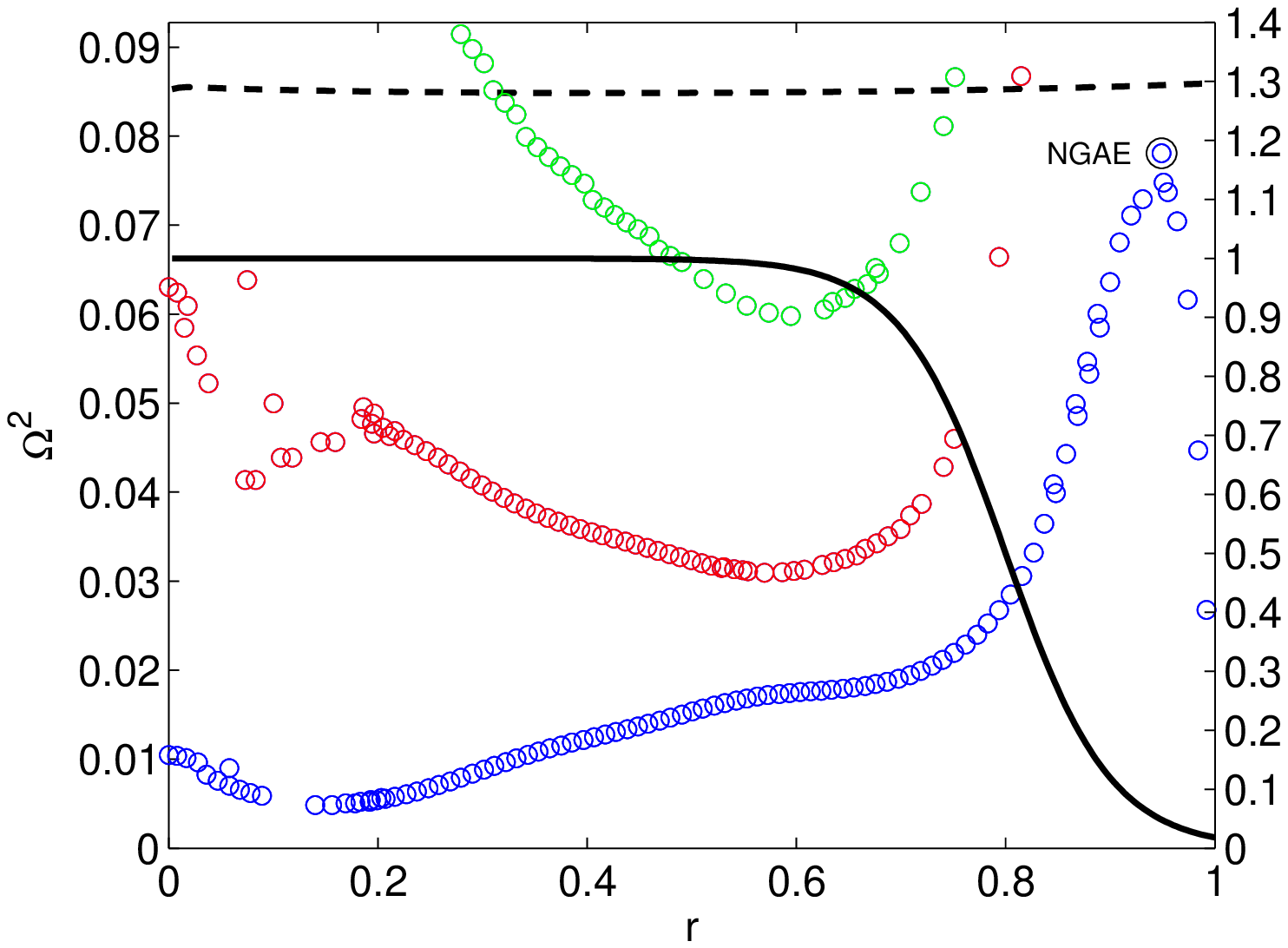}
\caption{\label{fig:H1_continuum}
Shear Alfv\'{e}n continuous spectrum as a function of radial position for the \HOneNF\ case computed using \CKA . The maximum of the $\left ( m = 10 , n = -13 \right )$ branch ($\textcolor{RoyalBlue}{\circ}$) near the edge of the plasma results in a \NGAE , which has its maximum at the location indicated. The $\left ( 8 , -10 \right )$ branch ($\textcolor{Red}{\circ}$) is also shown. The rotational transform profile, $\iota$, (dashed line) and normalised density profile, $\frac{n_i}{n_{i0}}$, (solid line) are plotted with the scale shown on the right axis.
}
\end{figure}

Finally, continuum damping was calculated for a non-conventional global Alfv\'{e}n eigenmode (\NGAE ) \cite{GAEs_and_NGAEs_in_stellarators} in an \HOneNF\ stellarator configuration \cite{H1_design_and_construction}. This device is an $N_{fp} = 3$ toroidal field period flexible heliac. The magnetic geometry of \HOneNF\ can be adjusted based on the ratio of currents through poloidal and toroidal field coils $\kappa_h$. Here we consider a case for which $\kappa_h = 0.33$. Again, the density profile used for the tokamak case described in Section~\ref{sec:tokamak} is used. The continuous spectrum obtained for the \HOneNF\ case is shown in figure~\ref{fig:H1_continuum}.

\begin{figure}[h] 
\centering 
\includegraphics[width=80mm]{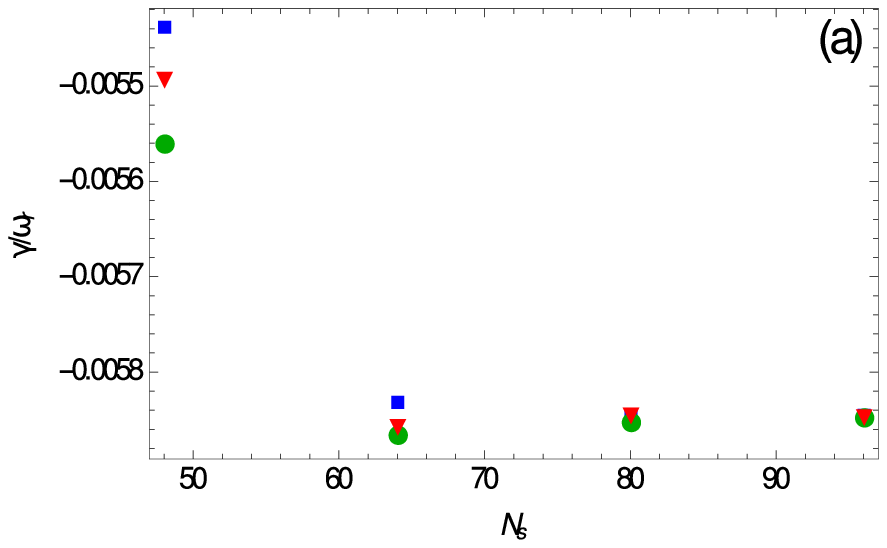}
\includegraphics[width=80mm]{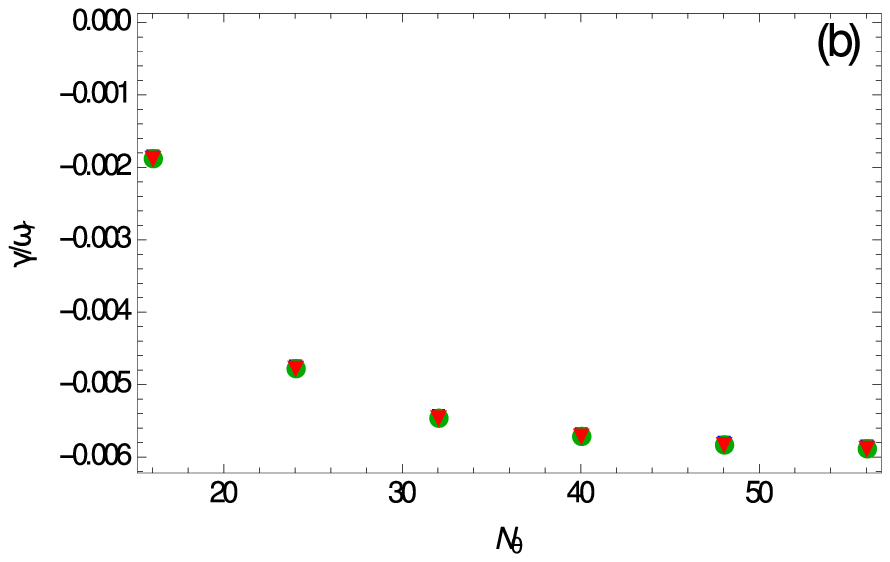}
\includegraphics[width=80mm]{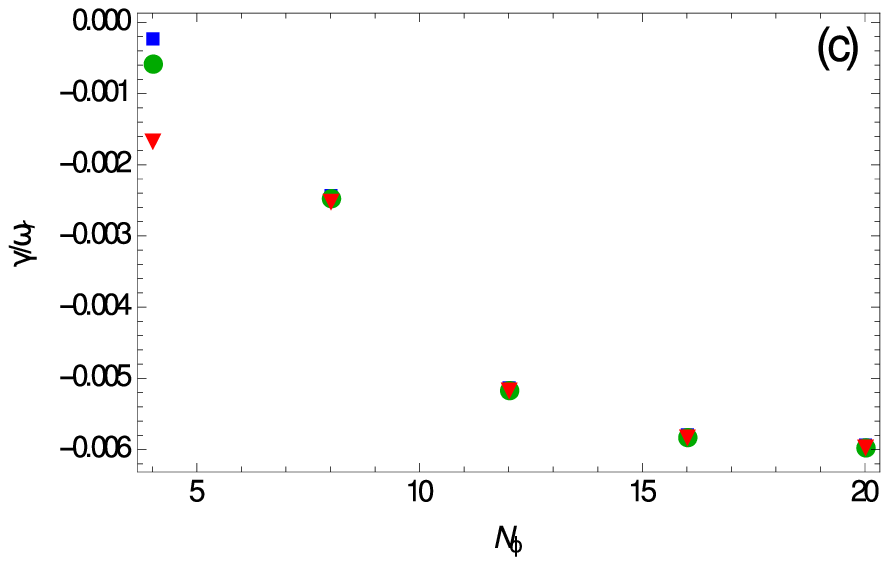}
\caption{\label{fig:H1_con}
Convergence of continuum damping ratio for an \NGAE\ due to the $\left ( 10 , -13 \right )$ shear Alfv\'{e}n wave harmonic in an \HOneNF\ plasma, with respect to (a) radial grid resolution $N_s$, (b) poloidal grid resolution $N_{\theta}$ and (c) toroidal grid resolution $N_{\phi}$. In each case, two of $N_s = 96$, $N_{\theta} = 48$ and $N_{\phi} = 16$ are fixed and the other parameter is varied. Complex contours are chosen with parameters $x_\beta = 0.1$ and $x_\gamma = 0.8$. Damping ratio is plotted for $\alpha = 0.1$ ($\textcolor{RoyalBlue}{\blacksquare}$), $\alpha = 0.2$ ($\textcolor{OliveGreen}{\bullet}$) and $\alpha = 0.5$ ($\textcolor{Red}{\blacktriangledown}$).
}
\end{figure}

An \NGAE\ is found which predominantly comprises the $\left ( 10 , -13 \right )$ harmonic. This type of mode is associated with a local maximum of the corresponding branch of the continuous spectrum. In this case the mode is localised near the edge of the plasma, where density declines sharply and $\iota$ approaches the $\frac{13}{10}$ rational surface. The \NGAE\ is found to be resonant with the $\left ( 8 , -10 \right )$ branch of the continuum at $s \approx 0.81$. The convergence of the continuum damping due to this resonance is shown in figure~\ref{fig:H1_con}. It is found that the normalised complex frequency of the mode is $\Omega = 0.277665 - 0.00162422 i$, corresponding to a damping ratio of $\frac{\gamma}{\omega_r} = 0.006647862$. As for the \WSX\ \TAE\ case, a relatively large poloidal and toroidal resolution is required due to the range of couplings induced by the complicated magnetic geometry.

\section{Conclusion}
Continuum damping of Alfv\'{e}n eigenmodes in complicated three-dimensional geometries can be calculated by solving the ideal \MHD\ eigenvalue problem along a complex contour in $s$. This is achieved by accurately fitting analytic functions to numerical equilibrium data and analytically continuing these to find values for complex $s$. An appropriate choice of complex contour allows relatively rapid convergence of this damping with respect to radial grid resolution. Coupling to numerous poloidal and toroidal harmonics can affect the damping of gap modes such as \TAEs\ and \EAEs\ in stellarators, as demonstrated by the convergence of damping with respect to grid resolution in the poloidal and toroidal direction.

Analytic continuation of the eigenfunctions computed along complex contours could be used to determine the structure of modes with continuum resonances for real values of $s$. These resonances are associated with sharp peaks in the amplitude of $\Phi$ and sudden changes in its phase. Conventional ideal \MHD\ codes find a sharp peak at the resonance but do not reproduce the jump condition at this singularity which can be found by complex contour integration. Properly incorporating the effect of continuum resonances on the mode structure would enable more accurate computation of fast particle drive and transport by incorporating the potentially significant effect of continuum resonances.

\section*{Acknowledgments}
The authors would like to thank the support of Australian Research Council grant FT0991899 and a grant supplied by the Group of Eight Australia / German Academic Exchange Service (DAAD). We would also like to acknowledge the assistance provided by Dr. B. Seiwald.



\end{document}